\documentclass[acmsmall,screen,nonacm]{acmart}

\usepackage{url}
\usepackage{booktabs}
\usepackage{tabularx}
\usepackage{multirow}
\usepackage{array}
\usepackage{bm}
\usepackage{cleveref}
\crefname{section}{§}{§§}
\Crefname{section}{§}{§§}
\usepackage{caption}
\usepackage{enumitem}
\setlist[itemize,1]{left=0pt}

\usepackage[linesnumbered,ruled,vlined]{algorithm2e}
\SetAlFnt{\small}
\SetAlCapFnt{\small}

\newcommand{\customsize}{\fontsize{7}{8}\selectfont}

\usepackage{listings, xcolor}
\definecolor{lightgreen}{rgb}{.12,.82,.31}

\lstdefinelanguage{Solidity}{
	keywords=[1]{anonymous, assembly, assert, balance, break, call, callcode, case, catch, class, constant, continue, constructor, contract, debugger, default, delegatecall, delete, do, else, emit, event, experimental, export, external, false, finally, for, function, gas, if, implements, import, in, indexed, instanceof, interface, internal, is, length, library, log0, log1, log2, log3, log4, memory, modifier, new, payable, pragma, private, protected, public, pure, push, require, return, returns, revert, selfdestruct, send, Solidity, storage, struct, suicide, super, switch, then, this, throw, transfer, true, try, typeof, using, value, view, while, with, addmod, ecrecover, keccak256, mulmod, ripemd160, sha256, sha3}, % generic keywords including crypto operations
	keywordstyle=[1]\color{blue}\bfseries,
	keywords=[2]{address, bool, byte, bytes, bytes1, bytes2, bytes3, bytes4, bytes5, bytes6, bytes7, bytes8, bytes9, bytes10, bytes11, bytes12, bytes13, bytes14, bytes15, bytes16, bytes17, bytes18, bytes19, bytes20, bytes21, bytes22, bytes23, bytes24, bytes25, bytes26, bytes27, bytes28, bytes29, bytes30, bytes31, bytes32, enum, int, int8, int16, int24, int32, int40, int48, int56, int64, int72, int80, int88, int96, int104, int112, int120, int128, int136, int144, int152, int160, int168, int176, int184, int192, int200, int208, int216, int224, int232, int240, int248, int256, mapping, string, uint, uint8, uint16, uint24, uint32, uint40, uint48, uint56, uint64, uint72, uint80, uint88, uint96, uint104, uint112, uint120, uint128, uint136, uint144, uint152, uint160, uint168, uint176, uint184, uint192, uint200, uint208, uint216, uint224, uint232, uint240, uint248, uint256, var, void, ether, finney, szabo, wei, days, hours, minutes, seconds, weeks, years},	% types; money and time units
	keywordstyle=[2]\color{teal}\bfseries,
	keywords=[3]{block, blockhash, coinbase, difficulty, gaslimit, number, timestamp, msg, data, gas, sender, sig, value, now, tx, gasprice, origin},	% environment variables
	keywordstyle=[3]\color{violet}\bfseries,
	identifierstyle=\color{black},
	sensitive=true,
	comment=[l]{//},
	morecomment=[s]{/*}{*/},
	commentstyle=\color{gray}\ttfamily,
	stringstyle=\color{lightgreen}\ttfamily,
	morestring=[b]',
	morestring=[b]"
}

\begin{document}

\title{NFTDELTA: Detecting Permission Control Vulnerabilities in NFT Contracts through Multi-View Learning}

\author{Hailu Kuang}
\affiliation{%
  \institution{Hainan University}
  \city{Haikou}
  \country{China}}
\email{hailukuang@hainanu.edu.cn}

\author{Xiaoqi Li}
\affiliation{%
  \institution{Hainan University}
  \city{Haikou}
  \country{China}}
\email{csxqli@ieee.org}

\author{Wenkai Li}
\affiliation{%
  \institution{Hainan University}
  \city{Haikou}
  \country{China}}
\email{cswkli@hainanu.edu.cn}

\author{Zongwei Li}
\affiliation{%
  \institution{Hainan University}
  \city{Haikou}
  \country{China}}
\email{lizw1017@hainanu.edu.cn}

\renewcommand{\shortauthors}{Kuang et al.}

\begin{abstract}

    Permission control vulnerabilities in Non-fungible token (NFT) contracts can result in significant financial losses, as attackers may exploit these weaknesses to gain unauthorized access or circumvent critical permission checks.
    In this paper, we propose NFTDELTA, a framework that leverages static analysis and multi-view learning to detect permission control vulnerabilities in NFT contracts. Specifically, we extract comprehensive function Control Flow Graph (CFG) information via two views: sequence features (representing execution paths) and graph features (capturing structural control flow). These two views are then integrated to create a unified code representation. We also define three specific categories of permission control vulnerabilities and employ a custom detector to identify defects through multi-view feature similarity analysis.
    Our evaluation of 795 popular NFT collections identified 241 confirmed permission control vulnerabilities, comprising 214 cases of Bypass Auth Reentrancy, 15 of Weak Auth Validation, and 12 of Loose Permission Management. Manual verification demonstrates the detector's high reliability, achieving an average precision of 97.92\% and an F1-score of 81.09\%. Furthermore, NFTDELTA demonstrates enhanced efficiency and scalability, proving its effectiveness in securing NFT ecosystems.
\end{abstract}


\maketitle

\section{INTRODUCTION}

    NFTs have emerged as an influential class of digital assets recently \cite{ante2023non} \cite{wang2021non} \cite{nadini2021mapping} \cite{ante2022non}.
    These unique digital assets represent ownership of specific items or content , distinguishing them from traditional fungible tokens (e.g., Ethereum), which are interchangeable or identical \cite{ante2022non} \cite{nadini2021mapping}  \cite{ethereum2024} \cite{sriman2022decentralized} \cite{li2025beyond}.
    The expansion of NFTs has catalyzed the advancement of NFT smart contracts. In contrast to conventional token contracts, NFT smart contracts are designed to comply with specific non-fungible token standards, such as ERC-721 \cite{ERC-721} or ERC-1155 \cite{ERC-1155}.
    The NFT contract serves as a specialized variation of a fungible token contract, focusing on key functions such as \texttt{mint}, \texttt{transfer}, \texttt{approve}, and \texttt{burn}, which are critical to the lifecycle of NFTs \cite{das2022understanding}. 
    
    Vulnerabilities in NFT contracts can lead to substantial financial losses \cite{wu2025security} \cite{yang2025multi} \cite{zhu2024sybil} \cite{zhang2025penetration}. For instance, in December 2023, a hacker exploited a reentrancy vulnerability in the NFT Trader project's contracts, resulting in the theft of high-value NFTs estimated at around 3 million dollars \cite{nftstolen2023}. Significant efforts have been dedicated to enhancing the security of NFT contracts \cite{yang2023definition} \cite{peng2026trifortis}. 
    The research by Yang et al.~\cite{yang2023definition} defined five common defects in NFT contracts by analyzing Q\&A site posts and security reports. These defects have the potential to cause significant financial losses. The researchers also used six popular vulnerability analysis tools to detect these defects. However, the results revealed that only the reentrancy defect was identifiable among the five defined defects. Moreover, none of these tools could detect the ERC-721 reentrancy attack in NFT contracts. To address this problem, the researchers proposed a precise symbolic execution tool, NFTGuard, that shows promising results in detecting defects in NFT contracts.
    
    However, challenges still exist in detecting vulnerabilities in NFT contracts:
    
    \noindent\textbf{Challenge 1 (C1): Vulnerability pattern coverage.}
    New smart contract vulnerabilities are discovered daily, and relying solely on predefined defect patterns for detection limits the ability to identify unknown defects \cite{chu2023survey}. 
    Permission control vulnerabilities in NFT contracts are critical. Proper permission control prevents unauthorized minting and destruction, ensures the legality of transfer operations, maintains the integrity of contract metadata, and prevents financial losses caused by malicious behavior \cite{liang2024identity} \cite{liu2022finding} \cite{liu2020smacs} \cite{zhong2024prettysmart}.
    
    \noindent\textbf{Challenge 2 (C2): Detection efficiency.}
    Symbol execution \cite{mythril2024} \cite{luu2016making} \cite{mossberg2019manticore}, static analysis \cite{feist2019slither}, and fuzzing test \cite{groce2021echidna} tools are commonly used to identify smart contract defects.
    However, these tools consume significant computing resources and can handle only a few contracts, given the large number of NFT contracts created from the growing NFT market \cite{white2022characterizing}. These limitations have resulted in the inability of such tools to be applied efficiently to large datasets.

    \noindent\textbf{Challenge 3 (C3): Feature extraction.}
    How to extract information from contract data to restore defect features is an indispensable topic for vulnerability detection. Typically, we build a CFG to capture the control flow of the code and an abstract syntax tree (AST) to obtain its syntactic structure. Some methods integrate both views to extract features associated with vulnerabilities \cite{neamtiu2005understanding} \cite{mythril2024} \cite{bu2025smartbugbert}. However, the accurate and efficient extraction of helpful information from contracts remains the focus of future research.

    \textbf{Our solution.}
    To address these challenges, we propose NFTDELTA, a generalized framework that enables customized vulnerability detection and determines the presence of vulnerabilities in a contract by checking vector similarity.

    \noindent\textbf{Solution to C1:}
    Unlike traditional vulnerability detection tools, NFTDELTA functions primarily as an auxiliary analysis framework. Instead of performing direct symbolic execution or taint analysis, it extracts features of vulnerable functions identified in detection reports and compares them with features of target functions to detect vulnerabilities. This design enables NFTDELTA to be compatible with most existing vulnerability detection tools and even manual audit reports.
    This study presents a practical solution for detecting permission control vulnerabilities in NFT contracts. We customize the detector interface to implement detection logic tailored to permission control scenarios in NFT contracts, enabling efficient and accurate vulnerability identification.

    \noindent\textbf{Solution to C2:}

    To improve detection efficiency, we adopt a non-training, similarity-based approach \cite{zhang2024combining} \cite{gao2019smartembed} \cite{huang2021hunting} \cite{zhu2022bytecode} \cite{tian2022ethereum}.
    We intentionally avoided supervised training because our goal is to generate expressive, high-fidelity embeddings for similarity comparison. Therefore, we exclusively utilized sequence and graph encoders to capture contextual and structural information.
    Vector-based detection is lightweight and resource-efficient, especially during inference. Once built, the vector database can be easily shared and reused, making it accessible and extensible. 
    As a complementary or alternative approach to directly using detection tools, this method utilizes high-quality embeddings for similarity-based detection, thereby reducing time consumption.

    \noindent\textbf{Solution to C3:}
    NFTDELTA fuse sequence and graph features extracted from function CFGs to construct a holistic feature representation for defective snippets. It extracts executable paths from CFGs, which then serve as inputs to the sequence learning model to derive contextual representations. Graph features are equally important as the other view for vulnerability analysis. We propose a novel approach to integrating sequence and graph features by leveraging an efficient graph representation learning model to capture intricate graph structures.

   The main contributions of this paper are as follows:
    \begin{itemize}
        \item We propose NFTDELTA, to the best of our knowledge, the first similarity-based NFT contract vulnerability detection framework that combines static analysis with multi-view feature learning through an innovative fusion strategy.
        \item We define three novel permission control vulnerabilities in NFT contracts: Weak Authorization Validation, Loose Permission Management, and Bypass Authorization Reentrancy. We further demonstrate that NFTDELTA can effectively detect these vulnerabilities in real-world contracts.
        \item Through evaluation, NFTDELTA outperforms other contract detection frameworks regarding throughput and scalability. We have uploaded the code, experimental results, and documentation in the paper to ensure reproducibility.
    \end{itemize}

    This paper is organized as follows:  
    \cref{sec::bg} provides an overview of non-fungible tokens and discusses the various token standards.
    \cref{sec::vulnerability} identifies and categorizes permission control vulnerabilities specific to NFT smart contracts, supplemented with illustrative examples.
    \cref{sec::NFTDELTA} presents the design and implementation of the NFTDELTA framework.
    \cref{sec::data_collection} outlines our methodology for collecting and filtering NFT contracts to construct the dataset.
    Subsequently, \cref{sec::exp} evaluates NFTDELTA's performance through comprehensive experiments.
    We then review relevant literature in \cref{sec::related}.
    Finally, we summarize our findings in \cref{sec::conclusion}.

\section{BACKGROUND}
\label{sec::bg}
    Ethereum \cite{ethereum2024} is an open-source, decentralized computing platform. It serves as the primary infrastructure for NFTs, utilizing immutable smart contracts written in Solidity to manage digital ownership \cite{solidity2024} . Unlike cryptocurrencies, NFTs represent unique assets, ranging from digital art to real-world collectibles, and adhere to standards such as ERC-721 \cite{ERC-721} or ERC-1155 \cite{ERC-1155}. These standards ensure interoperability across marketplaces like OpenSea and Blur \cite{opensea2024, blur2024} \cite{wu2025exploring}.
    The NFT lifecycle comprises three critical phases: minting, circulation, and destruction. During minting, creators deploy contracts to generate tokens and bind metadata. Once in circulation, ownership is transferred through sales or auctions, necessitating rigorous permission checks to prevent unauthorized transfers. Finally, destruction (burning) involves permanently sending tokens to a zero address. Since these operations irreversibly alter asset states, the security of the underlying permission logic is critical \cite{ante2023non}.

    NFT smart contracts predominantly adhere to two Ethereum standards: ERC-721 \cite{ERC-721} and ERC-1155 \cite{ERC-1155}. ERC-721 establishes the foundational interfaces for non-fungible assets, mandating strict logic for ownership tracking and access control. Critical functions include \texttt{ownerOf} for identity verification, \texttt{safeTransferFrom} for asset transfer, and \texttt{setApprovalForAll} for delegating management rights to third-party operators. Each asset is distinguished by a unique \texttt{\_tokenId}. ERC-1155 extends this capability by allowing a single contract to manage mixed asset types (fungible and non-fungible). It optimizes efficiency through batch transfers and supports attaching auxiliary data to transactions. While offering flexibility, these complex interaction patterns introduce additional challenges for permission validation.

\section{PERMISSION CONTROL VULNERABILITY}
\label{sec::vulnerability}

    This section will explore permission control vulnerabilities in NFT smart contracts. We will define three types of vulnerabilities and identify potential exploitations by providing practical examples.
    A permission control vulnerability arises from a flaw in a system's design or implementation. It allows an unauthorized user or entity to execute an operation that should only be performed by a specific privileged holder. Exploiting these vulnerabilities enables attackers to gain elevated privileges, potentially leading to serious issues such as the theft of user assets and malicious contract abuse. 
 
    Permission control vulnerabilities in NFT smart contracts can be classified into three main groups: Weak Authorization Validation, Loose Permission Management, and Bypass Authorization Reentrancy. Detailed definitions are in Table \ref{tab:permission_control_vulnerability}.
    Notably, any vulnerability that meets the criterion above falls under the permission control category, including defects not covered in the current classification. Our definition is intended as an extensible foundation, and we plan to expand it as more patterns are discovered.

    \begin{table}[ht]
    \small
    \centering
    \caption{Categories of Permission Control Vulnerabilities and Corresponding Defect Types.}
    \label{tab:permission_control_vulnerability}
    \begin{tabular}{>{\centering\arraybackslash}m{2cm} >{\centering\arraybackslash}m{2cm} >{\raggedright\arraybackslash}m{6cm}}
        \toprule
        \textbf{Vulnerability} & \textbf{Type} & \textbf{Description}\\
        \midrule
        \multirow{2}{1.6cm}{\centering Weak Auth Validation} & Standard & Failed to follow permission requirements specified in standard.\\
        & Non-standard & Failed to follow development requirements not explicitly mentioned in the standard.\\
        \midrule
        \multirow{2}{1.6cm}{\centering Loose Perm Management} & Assign & Wrong permission setting in functions.\\
        & Revoke & Authorizations not revoked correctly in functions.\\
        \midrule
        Bypass Auth & Reentrancy & Reentrancy that possibly bypass permission verification.\\
        \bottomrule
    \end{tabular}
    \end{table}

    \subsection{Weak Authorization Validation}
        We consider contracts that fail to meet the permission requirements specified in the token standard or fail to follow the development practices not explicitly mentioned in the standard to have Weak Authorization Validation vulnerabilities.
        
    \textbf{Standard:}
        The NFT token standard incorporates authorization validation to ensure secure operations. In the ERC-721 standard \cite{ERC-721}, for example, the standard specifies that the \texttt{transferFrom} function must authenticate the caller's identity to be the current owner, an authorized operator, or the approved address for this NFT before executing ownership changes. In most implications, authorization validation is typically enforced through the \texttt{require} statements or utilizes \texttt{if-revert} patterns as a gas-efficient alternative.

    \textbf{Non-standard:}
        The NFT token standard does not enforce specific permission checks for functions such as \texttt{mint} and \texttt{burn}. It does not mandate implementation for custom functions, modifiers, or external integrations such as role-based permission control contracts. This implementation avoids overly detailed standards restricting developers' flexibility when designing contracts. However, greater flexibility also entails greater risk. Developers may introduce more possibilities and uncertainties as they make all the decisions. 
        
        \textbf{Example:} In the \texttt{transferFrom} function of Listing \ref{lst:transferFrom}, the function \texttt{\_isApprovedOrOwner} in the require statement is used to validate the identity of the function caller. However, there is an imperceptible error in this validation function. In line 8 of listing \ref{lst:transferFrom}, the function \texttt{getApproved} returns the result of the authorized transferor of the specified NFT. It should have been compared to the function caller, but instead, it is compared to the token owner. As a result, the function always returns true, and the \texttt{require} statement loses its effectiveness in validating the identity.

        \begin{lstlisting}[label=lst:transferFrom, caption={Example of the transferFrom Function and its Internal Functions.}]
function transferFrom(address from, address to, uint256 tokenId) public virtual  {
    require(_isApprovedOrOwner(msgSender(), tokenId));
    _transfer(from, to, tokenId);}

function _isApprovedOrOwner(address spender, uint256 tokenId) internal view virtual returns (bool) {
    require(_exists(tokenId));
    address owner = ERC721.ownerOf(tokenId);
    return (spender == owner || getApproved(tokenId) == owner || isApprovedForAll(owner, spender));}

function _transfer(address from, address to, uint256 tokenId) internal virtual {
    require(ERC721.ownerOf(tokenId) == from);
    require(to != address(0));
    _approve(address(0), tokenId);
    _balances[from] -= 1;
    _balances[to] += 1;
    _owners[tokenId] += to;
    emit Transfer(from, to, tokenId);}
        \end{lstlisting}

        \subsection{Loose Permission Management}
        We refer to contracts that include improper permission settings or revocation operations that result in unintended state changes as having Loose Permission Management.
        
        \textbf{Assignment:}
        In contrast to authorization validation, which verifies the caller's identity, permission management focuses on accurately setting and revoking permissions to prevent malicious behavior. In NFT contracts, modifications to certain state variables can directly affect account-level permission management. For example, the state variable \texttt{\_tokenApprovals} records the approved address for a token identified by \texttt{tokenId}. Once authorized, this approved address is permitted to transfer the corresponding token on behalf of the token owner. Within the \texttt{\_approve} function, the caller—typically the token owner—specifies the approved address \texttt{to} along with the \texttt{tokenId} to update the state variable, thereby granting transfer authorization for that owned token.

        \textbf{Revocation:}
        Likewise, permission revocation during NFT transfers or burns is typically handled by clearing privileges through updates to the state variable. Revocation is commonly implemented by resetting the relevant variables to the zero address (\texttt{0x00}) or explicitly removing them using the \texttt{delete} statement. In practice, these operations are performed on variables such as \texttt{\_tokenApprovals} and \texttt{\_owners} within the \texttt{\_transfer} and \texttt{burn} functions to revoke the approved address and ownership mappings accordingly.
        
        \textbf{Example:} In the \texttt{\_transfer} function of Listing \ref{lst:transferFrom}, while the \texttt{\_approve} function successfully removes all previously assigned approvals and handles the account balance correctly, it fails to revoke ownership from original owners when transferring in line 16. As a result, the previous token holder may still retain ownership of the NFT. It is worth mentioning that while some tokens are designed to allow multiple owners to share the same token, this practice is not recommended as it may lead to status derailment and confusion.
        
        \subsection{Bypass Authorization Reentrancy}

        A Bypass Authorization Reentry represents a contract vulnerability that enables unauthorized access to privileged operations through recursive external calls that circumvent proper authorization validation. In NFT contracts, the \texttt{\_checkOnERC721Received} function manifests to verify ERC-721 compatibility of the recipient address \texttt{to}. While this function initiates a callback to the recipient's \texttt{onERC721Received} method to validate token receipt capability, it creates a potential attack vector. A malicious address could exploit this by implementing a manipulated \texttt{onERC721Received} function that recursively calls the original function, bypassing critical permission checks and operating the contract's state transition.
    
        \textbf{Example:} The mint function in Listing \ref{lst:mint} is vulnerable to a Bypass Authorization Reentrancy vulnerability. In this function, the designer did not follow the check-effect-interaction pattern, allowing an attacker to rewrite the \texttt{onERC721Received} function and repeatedly call the \texttt{mintNFT} function. As a result, the attacker can bypass the \texttt{\_isMintable} check in line 2 and mint more NFTs than allowed, devaluing the NFTs' market capitalization.

        \begin{lstlisting}[label=lst:mint, caption={Example of the mint Function.}]
function mint(uint256 nftType, uint256 count, address to) public onlyValidSender{
    require(!_isMintable(nftType));
    // Variable initialization
    for (uint256 tokenId = tokenNumber; tokenId <= tokenLimit; tokenId++) {
        _owners[tokenId] = to;
        require(_checkOnERC721Received(address(0), to, tokenId, _data));}
    // State variable updates
    emit ConsecutiveTransfer(tokenNumber, tokenLimit, address(0), to);}
        \end{lstlisting}

\section{NFTDELTA}
\label{sec::NFTDELTA}

    In this study, we propose NFTDELTA, a generalized framework for customized vulnerability detection in NFT contracts. NFTDELTA identifies vulnerabilities by leveraging multi-view embeddings and vector similarity analysis. The overall workflow of NFTDELTA is illustrated in Figure \ref{fig:NFTDELTA_workflow}. As shown in the figure, the data processing pipeline of our framework comprises three main phases, which are further decomposed into the following six concrete steps:

    \begin{figure*}[htbp]
        \centering
        \includegraphics[width=\textwidth, trim=2 22 2 2, clip]{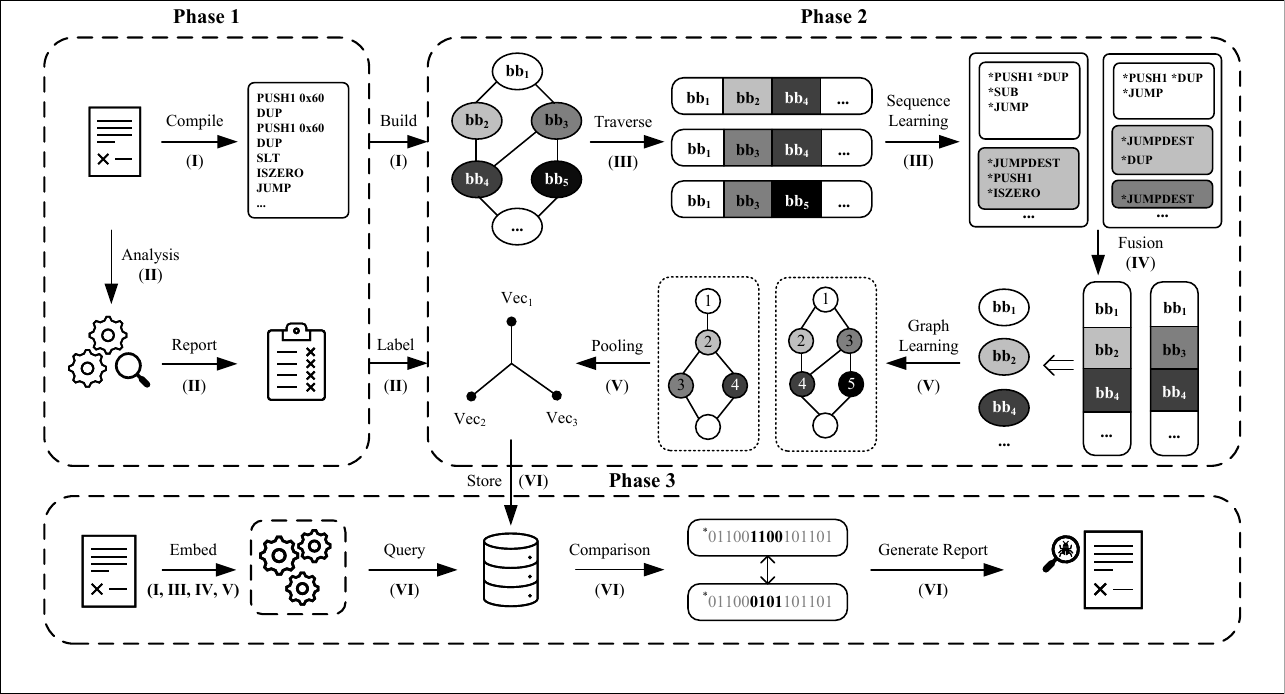}
        \caption{The Overall Workflow of the NFTDELTA Framework, Comprising Three Phases: (1) Static Analysis and Defect Mapping, (2) Feature Extraction with Multi‑View Fusion, and (3) Feature Storage with Similarity Detection. These Phases are Further Decomposed into Six Steps: Compilation and Preprocessing, Static Analysis, Sequence Feature Extraction, Reorganization and Fusion, Graph Feature Extraction, and Detection.}
        \label{fig:NFTDELTA_workflow}
        \vspace{-3ex}
    \end{figure*}

        \textit{(\uppercase\expandafter{\romannumeral1}) Compile and Preprocess.}
        Our method begins by compiling the NFT contract into bytecode and constructing function-level CFGs. Bytecode serves as a low-level representation of contract function execution instructions, from which corresponding opcodes can be retrieved through a deterministic mapping \cite{pasqua2023enhancing}. Opcodes $o_j \in \mathcal{O}$, known as Ethereum Virtual Machine (EVM) instructions, represent the set of operations necessary for executing smart contracts \cite{evmopcode2024}.
        Function-level CFGs $G = (\mathcal{B}, \mathcal{E})$ are constructed by analyzing the jump and branch semantics within opcode sequences. Each node, or basic block $b_i \in \mathcal{B}$, is defined as a subsequence of opcodes:

        \begin{equation}
            b_i = [o_1, o_2, \dots, o_{k_i}], \quad o_j \in \mathcal{O} 
        \end{equation}

        Each directed edge $e = (b_s, b_t) \in \mathcal{E}$ represents a control flow transition from basic block $b_s$ to $b_t$. This graph structure enables traversal algorithms such as loop-avoiding Depth-first search (DFS) to extract a set of feasible execution paths $\{p_1, p_2, \dots, p_n\} \subseteq \mathcal{P}$, where each path $p_i$ is an ordered sequence of basic blocks:

        \begin{equation}
        p_i = [b_1, b_2, \dots, b_{x_i}], \quad b_j \in \mathcal{B} 
        \end{equation}

        \textit{(\uppercase\expandafter{\romannumeral2}) Static Analysis.}
        Static analysis is an efficient technique for evaluating programs without execution \cite{ghaleb2020effective}. 
        In our study, we first parse the smart contract source code into an AST and transform it into a linear Intermediate Representation (IR) based on Static Single Assignment (SSA) form. This structural transformation facilitates the construction of CFG and the extraction of precise definition-use chains for variable tracking. By leveraging taint analysis to propagate data dependencies across execution paths and applying specialized heuristic constraints tailored to NFT business logic, we effectively identify permission control vulnerabilities \cite{feist2019slither}.

        The Weak Authorization Validation and Loose Permission Management detectors employ a heuristic pattern-matching approach. These detectors begin by identifying public functions associated with permission control as entry points and traversing them, along with their internal calls, using a breadth-first search (BFS). Each traversed function checks whether the parameters match predefined patterns, including parameter count, data types, and specific modifiers, to filter out irrelevant cases. For target functions, it merges the IR operations of the function and its modifiers to determine whether the logic deviates from expected permission control patterns.
        Meanwhile, the Bypass Authorization Reentrancy detector identifies vulnerabilities related to external calls and modifications to state variables. It examines the function's call sequence to detect external calls that deviate from the current execution flow, and it tracks variables read before an external call that are not immediately written back.

\begin{figure}[htbp]
            \centering
            \includegraphics[width=0.6\textwidth, trim=20 20 10 20, clip]{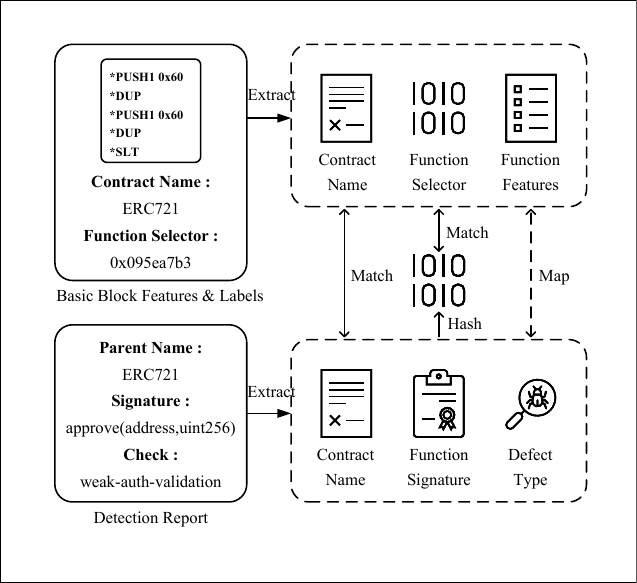}
            \caption{Mapping Defects and Features via Function Selectors.}
            \label{fig:defect_mapping}
        \end{figure}

        We also establish mappings between analysis reports and instruction-level features to extract vulnerable embeddings. To support scalability, defect mapping is achieved through a straightforward mechanism, enabling NFTDELTA to integrate seamlessly with various vulnerability detection tools. Thereby, enhance their effectiveness in identifying diverse types of vulnerabilities.
        As illustrated in Figure \ref{fig:defect_mapping}, our framework extracts basic block features from the contract's CFG and records their positions. These positions are then aligned with metadata from the detection report by hashing the reported function signature into a 4-byte selector using the Keccak-256 algorithm and comparing it with function selectors extracted from the CFG. In this way, NFTDELTA can accurately associate low-level opcode features with their corresponding vulnerability types.
        This mechanism enables seamless integration with detection tools, provided that the detection report includes the contract name, function signature, and the associated vulnerability type for each vulnerable function.

        \textit{(\uppercase\expandafter{\romannumeral3}) Sequence Feature Extraction.}
        We process the extracted EVM instruction sequences, derived by traversing execution paths of contract functions, in two steps. First, we apply a word embedding method, Word2Vec \cite{mikolov2013efficient}, to embed each opcode into a continuous vector space. Most EVM opcodes are composed of commonly seen tokens (e.g., CALLDATASIZE), which are generally covered even by the default vocabulary of Word2Vec models. In rare cases (e.g., unexpected or corrupted characters), we assign zero vectors to the opcode features to prevent runtime errors. We define an embedding function $\phi : \mathcal{O} \rightarrow \mathbb{R}^d$ that transforms each opcode $o_j \in \mathcal{O}$ into:

        \begin{equation} 
            \mathbf{v}_j = \phi(o_j), \quad \mathbf{v}_j \in \mathbb{R}^d 
        \end{equation}
        
        For a given execution path $p = [o_1, o_2, \dots, o_m]$, the embedded sequence is represented as (where $[\cdot;\cdot]$ denotes vertical stacking):

        \begin{equation}
        \mathbf{V}_p = [\phi(o_1); \phi(o_2); \dots; \phi(o_m)] \in \mathbb{R}^{m \times d} 
        \end{equation}

        Subsequently, to capture contextual and structural dependencies within the execution paths of each function, we feed the embedded sequences $\{ \mathbf{V}_{p_1}, \mathbf{V}_{p_2}, \dots, \mathbf{V}_{p_n}\}$ into a sequence learning model $\mathcal{F}$ (Performer \cite{choromanski2020rethinking}).
        To facilitate batch processing, each $\mathbf{V}_{p_j}$ is padded to a fixed maximum path length $m_{\text{max}}$. The resulting input tensor is:

        \begin{equation} 
        \mathbf{H}_i = \mathcal{F}([\mathbf{V}_{p_1}, \mathbf{V}_{p_2}, \dots, \mathbf{V}_{p_n}]) \in \mathbb{R}^{n \times m_{\text{max}} \times d}
        \end{equation}

        Here, $\mathbf{H}_i$ denotes the function-level representation that captures fine-grained opcode semantics and global control-flow structure, enabling more precise vulnerability detection.

\begin{figure}[htbp]
            \centering
            \includegraphics[width=0.6\textwidth, trim=24 14 2 20, clip]{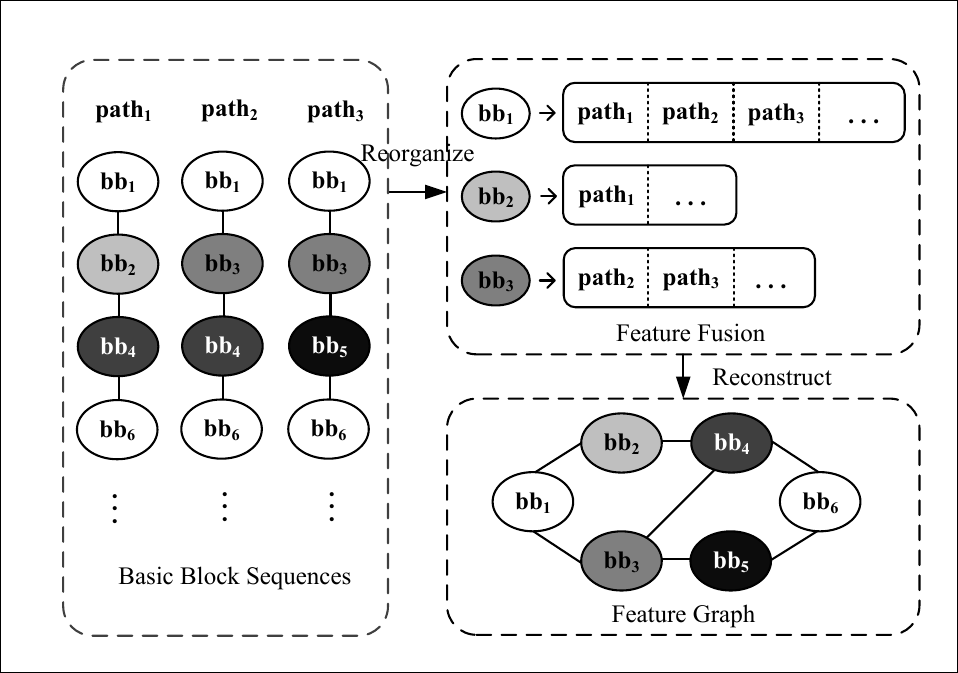}
            \caption{Reorganize and Fusion Sequence Features into Graph Features.}
            \label{fig:remapping_and_fusion}
        \end{figure}

        \textit{(\uppercase\expandafter{\romannumeral4}) Reorganization and Fusion.}
        To effectively capture the graph structure of the CFG and enhance feature semantics, we reorganize the embedded sequence features based on the CFG topology. Since executable paths are extracted by traversing the CFG, the same basic block may appear in multiple execution paths and at varying positions. We apply a feature fusion method that aggregates contextual embeddings from all relevant paths to assign each basic block a unique and consistent representation, as illustrated in Figure \ref{fig:remapping_and_fusion}.

        Let $b_j$ denote a basic block in the CFG, with each $b_j$ corresponding to a fixed dimension $k_j$ (e.g., the number of instructions). Let $\mathcal{P}(b_j)$ be the set of execution paths in which $b_j$ appears. For each occurrence of $b_j$ in path $p_i \in \mathcal{P}(b_j)$, we extract the contextual representation $\mathbf{h}_{p_i}^{(b_j)}$ from the path-level embedding output $\mathbf{H}_i$. The final representation of $b_j$ is obtained by weighted fusion:

        \begin{equation}
            \mathbf{Z}_j = \text{Fuse} \left( \left\{ \mathbf{h}_{p_i}^{(b_j)} \;\middle|\; p_i \in \mathcal{P}(b_j) \right\} \right) \in \mathbb{R}^{k_j \times d}
        \end{equation}

        Here, $\text{Fuse}(\cdot)$ denotes an attention-inspired aggregation function that assigns weights based on the relative position of $b_j$ in each path. The attention weight $w_i$ for $b_j$ in path $p_i$ is computed as:

        \begin{equation}\label{eqa:weights}
            w_i = \text{softmax}_{p_i \in \mathcal{P}(b_j)}\left(\exp\left(\alpha \cdot \text{clip}\left(\frac{s_i - s_{\min}}{\overline{L}}, 0, c\right)\right)\right)
        \end{equation}
        
        Where $s_i$ is the starting position of $b_j$ in path $p_i$, $s_{min}$ represents its minimum starting position. $\overline{L}$ refers to the average block length in the function. The hyperparameter $\alpha$ is a scaling factor (set to 0.6), and $c$ is a truncation limit (set to 5) used to prevent gradient explosion. 
        This fusion strategy prioritizes embeddings from relatively deeper positions in the execution path, as later blocks are often associated with more complex control flow and richer semantic context.

        \textit{(\uppercase\expandafter{\romannumeral5}) Graph Feature Extraction.}
        After contextual fusion, the instruction-level features of a function are reconstructed into a control-flow-reflective graph. Specifically, we treat each instruction as a node in a newly reconstructed graph $\mathcal{G} = (\mathcal{V}, \mathcal{E})$, where $\mathcal{V}$ denotes the set of instruction nodes, and $\mathcal{E}$ contains two types of directed edges: $\mathcal{E}_{\text{CFG}}$ representing control flow between basic block, and $\mathcal{E}_{\text{Seq}}$ capturing the sequential execution order of instructions. Each node $v_j \in \mathcal{V}$ is initialized with a contextual feature vector $\mathbf{z}_j$ 
        extracted from the block-level embedding $\mathbf{Z}_b \in \mathbb{R}^{k_j \times d}$.
        
        To obtain structurally expressive graph features, we employ FastGAT \cite{srinivasa2020fast}, a graph representation learning model designed for efficient global attention calculation, to encode the graph. Formally, node representations are updated as follows:

        \begin{equation}
        \mathbf{h}_j^{(l+1)} = \sigma \left( \sum_{u \in \mathcal{N}(j)} \alpha_{ju}^{(l)} \mathbf{W}^{(l)} \mathbf{h}_u^{(l)} \right),
        \end{equation}

        Here, $\mathbf{h}_j^{(l)}$ is the representation of node $v_j$ at layer $l$; $\mathcal{N}(j)$ is the neighborhood of $v_j$; $\alpha_{ju}^{(l)}$ are attention weights computed over edges (include $\mathcal{E}_{\text{CFG}}$ and $\mathcal{E}_{\text{Seq}}$); $\mathbf{W}^{(l)}$ is the layer-specific transformation matrix, and $\sigma(\cdot)$ is a non-linear activation.

        Finally, to mitigate the complexity introduced by fine-grained instruction representations, we employ an attention-pooling mechanism to aggregate instruction features within each basic block into a single high-level vector, which serves as the final function-level representation.
        Let $\mathbf{V}_{b_j} \in \mathbb{R}^{k_j \times d}$ be the instruction embeddings within basic block $b_j$. The block-level representation $\mathbf{z}_j$ is obtained through:

        \begin{equation}
        \mathbf{z}_j = \text{Attn}(\mathbf{V}_{b_j}) = \text{softmax}(\mathbf{a}^\top \tanh(\mathbf{W}\mathbf{V}_{b_j}^\top))\mathbf{V}_{b_j}
        \end{equation}
        
        Here, $\mathbf{W} \in \mathbb{R}^{d' \times d}$ and $\mathbf{a} \in \mathbb{R}^{d'}$ are learnable parameters. This mechanism highlights semantically salient instructions and preserves the function’s structure by encoding it as basic block embeddings.

        \textit{(\uppercase\expandafter{\romannumeral6}) Detection.} 
        The vulnerability detection process consists of two phases: embedding and detection.
        In the embedding phase, NFTDELTA utilizes embedding modules to extract features. Based on the analysis results from the detectors, it filters out defective function representations and stores them, along with their position labels, in a vector database, Hnswlib \cite{malkov2018efficient}, thereby forming a pre-collected vulnerability dataset.
        
        In the detection phase, we also perform feature embedding. However, the key difference is that the static analysis model is not required for detecting vulnerabilities in new contracts, thereby reducing computational overhead and improving detection efficiency. 
        Finally, we compare the extracted function features with those in the database by calculating the Euclidean distance between the vectors. We determine whether two functions are similar by applying multiple criteria, thereby identifying potential vulnerabilities.

\section{DATA COLLECTION}
    \label{sec::data_collection}
    
    In our study, we collected two datasets of NFT contracts—one for generating embedded features and the other for verifying the performance of our framework. The dataset source, acquisition methods, and usage of these datasets are shown in Table \ref{tab:dataset_collection}. Additionally, Table \ref{tab:dataset_features} presents key characteristics of the two datasets, including the number of contracts, average number of logical lines of code (excluding blank lines and comments), instructions, and functions.

    \begin{table}[ht]
    \small
    \centering
    \caption{Composition and Sources of Datasets for Embedding and Validation.}
    \label{tab:dataset_collection}
    \begin{tabular}{>{\centering\arraybackslash}m{2.6cm} >{\centering\arraybackslash}m{1.2cm} >{\centering\arraybackslash}m{3cm} >{\raggedright\arraybackslash}m{4cm} } % 使用 p{width} 设定列宽
        \toprule
        \textbf{Dataset} & \textbf{Source} & \textbf{Method} & \textbf{Description}\\
        \midrule
        Embedding Dataset & Pubilc & Compilability Filtering & Prepare for similarity detection\\
        \midrule
        Validation Dataset  & OpenSea & API Access \& Filtering & Validate detection accuracy\\
        \bottomrule
    \end{tabular}
    \end{table}

    As shown in Table \ref{tab:dataset_features}, both datasets exhibit high overall complexity; however, contracts sourced from OpenSea are more complex across multiple dimensions. This is because contracts deployed on OpenSea often implement more sophisticated business logic to support features such as marketplace transactions, royalty calculations, and multi-signature verification.

    \begin{table}[ht]
    \small
    \centering
    \caption{Statistics for Embedding and Validation Dataset.}
    \label{tab:dataset_features}
    \begin{tabular}{>{\centering\arraybackslash}m{1.2cm} >{\centering\arraybackslash}m{1.2cm} >{\centering\arraybackslash}m{1.2cm} >{\centering\arraybackslash}m{1.3cm} >{\centering\arraybackslash}m{1.3cm}} % 使用 p{width} 设定列宽
        \toprule
        \textbf{Dataset} & \textbf{\#Contracts} & \textbf{Avg. LOC} & \textbf{Avg. Instr} & \textbf{Avg. Func}\\
        \midrule
        Embedding & 3,146 & 655.5 & 14,250.1 & 30.7\\
        \midrule
        Validation & 795 & 795.9 & 16,306.0 & 51.2\\
        \bottomrule
    \end{tabular}
    \end{table}

    \textbf{Embedding dataset:} The dataset used for embedding was derived from the research conducted by Yang et al. \cite{yang2023definition}. It comprises 16,527 Ethereum smart contracts collected from an open-source smart contract repository \cite{smart_contract_sanctuary} as of August 2022. These contracts were filtered using the keywords such as ERC721 and NFT, and each was validated on Etherscan \cite{etherscan2024}. Compared to the Smart Bugs dataset \cite{FerreiraEtAl2020ASE}, this dataset exhibits more complex functional logic and better captures the current characteristics of NFT contracts.
    However, due to the constraints of the experimental environment and tool-specific limitations, we found that not all contracts could be compiled and executed successfully. To address this, we established an automated contract processing and validation workflow, leveraging tools such as truffle-flattener, solc-select, and crytic-compile to construct the verification script \cite{truffleflattener2024} \cite{solcselect2024} \cite{cryticcompile2024}.
    
    To simplify the compilation process and ensure data integrity, we utilized truffle-flattener to analyze inter-contract dependencies, consolidating multiple interrelated contract files into a single flattened contract. This ensured that all dependencies and structures were processed in the correct order.
    We used the solc-select tool to dynamically switch the compiler version during compilation based on each contract's version declaration. This not only preserved the original code semantics but also ensured consistent bytecode generation, facilitating CFG construction.
    Through automated validation, we identified 3,146 NFT smart contracts that comply with the ERC-721 standard and can be successfully flattened and compiled to construct the embedding dataset.

    \textbf{Validation dataset:} 
    We collected data samples for the validation dataset by accessing APIs, focusing on Ethereum contracts written in Solidity that comply with the ERC-721 standard. Through the OpenSea API \cite{openseaapi2024}, we gathered 3,178 NFT contract projects, sorted by their market capitalization on OpenSea, as of September 2024. This collection features well-known projects, including Bored Ape Yacht Club, Pudgy Penguins, and Autoglyphs. Each contract holds an average market value of hundreds of thousands of dollars and plays a significant role in practical applications.

    Next, we utilized the Etherscan API \cite{etherscan2024} to retrieve the source code for these contracts based on their addresses and filtered them using the automated process described earlier. After filtering, we identified 795 NFT smart contracts that comply with the ERC-721 standard and can be successfully compiled and executed.
    These smart contracts span a diverse range of Solidity compiler versions, from 0.4.23 to 0.8.26, and have been validated on Etherscan.

\section{EXPERIMENTAL EVALUATION}
\label{sec::exp}
In this section, we will conduct experiments on NFTDELTA to thoroughly evaluate its performance and address the following questions:

\noindent\textbf{\underline{RQ1 (Accuracy):}} How accurately does NFTDELTA detect three permission control vulnerabilities? What proportion of these vulnerabilities exists in real-world NFT contracts?\\
\textbf{\underline{RQ2 (Comparison):}} Can other detection tools identify permission control vulnerabilities in NFT contracts? What advantages does NFTDELTA offer compared to these frameworks?\\ 
\textbf{\underline{RQ3 (Ablation):}} How do the diverse modules and parameters of NFTDELTA influence the similarity detection results?

    \subsection*{Experimental Setup}
    \label{subsec::exp_setup}
    Our experiments are conducted on a server equipped with an Intel Core i7-13700KF CPU, 32GB of RAM, and an NVIDIA RTX 4070 GPU, running Ubuntu 24.04. The two datasets used in the experiment, comprising 3,146 contracts for embedding and 795 for validation, are described in \cref{sec::data_collection}.
    
    In our model, we set the dimensions of the word, sequence, graph, and basic block embeddings to 64, 96, and 128, respectively. For word embedding, the window size is set to 5. The sequence embedding has 6 layers and 8 multi-attention heads. The graph embedding consists of 3 layers, with 8 attention heads in the first two layers and one in the third. Note that to ensure the stability and reproducibility of the embedding results, we set a fixed global random seed value of 42 for the entire embedding process.
    In the database, we set the index \texttt{ef\_construction} 's search depth to 200 and the maximum number of neighbors per node to 16. We utilize Euclidean distance as the metric for measuring similarity.

    \subsection*{RQ1: Accuracy}
    
    To answer RQ1, we utilized NFTDELTA's static analysis detectors on the validation dataset to evaluate detection accuracy for permission control vulnerabilities. We manually reviewed each contract that reported a vulnerability, categorizing them into true positives (TP) and false positives (FP). Then, we calculated the precision of the detection results. If a contract was associated with multiple types of vulnerabilities, we repeated the calculation for each type and added it to the corresponding category. Two researchers with experience in smart contract auditing independently conducted the labeling. We followed guidelines based on our taxonomy of NFT permission control vulnerabilities. In cases of disagreement, a third expert adjudicator determined the final label. All labeling results are documented, and we will release the annotation records alongside the dataset to ensure transparency and reproducibility.
    
    The experimental results in Table \ref{tab:static_analysis_accuracy} outline the characteristics of the real-world validation dataset and the performance of the static analysis detector.
    It is crucial to distinguish between a theoretical defect and a practical security threat. Many flagged functions, especially in minting, are protected by privileged access controls (e.g., \texttt{onlyOwner} modifier), which prevent attackers from accessing them. Furthermore, modifications to certain state variables may be benign, causing no significant financial loss. Thus, the presence of these defects does not necessarily imply the contract is susceptible to malicious attacks.

    \begin{table}[ht]
    \small
    \centering
    \caption{Performance Evaluation of the NFTDELTA Static Analysis Model, Including the Counts of TP, FP, and FN Reports, with the Detector's Precision, Recall, and F1 Score in Identifying Each Defect Category.}
    \label{tab:static_analysis_accuracy}

    \begin{tabular}{>{\centering\arraybackslash}m{3.2cm} >{\centering\arraybackslash}m{0.8cm} >{\centering\arraybackslash}m{0.8cm} >{\centering\arraybackslash}m{0.8cm} > {\centering\arraybackslash}m{1cm} >
    {\centering\arraybackslash}m{1cm} >{\centering\arraybackslash}m{1cm}}
        \toprule
        \textbf{Defect Category} & \textbf{\#TP} & \textbf{\#FP} & \textbf{\#FN} & \textbf{Prec(\%)} & \textbf{Recall(\%)} & \textbf{F1(\%)}\\
        \midrule
        Weak Auth Validation & 15 & 1 & 8 & 93.75 & 65.22 & 76.92\\
        \midrule
        Loose Perm Management & 12 & 0 & 10 & 100 & 54.55 & 70.59\\
        \midrule
        Bypass Auth Reentrancy & 214 & 0 & 19 & 100 & 91.85 & 95.76\\
        \bottomrule
    \end{tabular}
    \end{table}
    
    The experimental results demonstrate that the NFTDELTA detector is highly effective in identifying permission control vulnerabilities, achieving an average precision of 97.92\%. Bypass Auth Reentrancy represents the most prevalent and accurately detected category. The detector identified 214 True Positives with 100\% precision and a high recall of 91.85\%, resulting in an F1-score of 95.76\%. 
    These vulnerabilities are primarily located in \texttt{mint} functions, where interactions with external contracts via the \texttt{\_checkOnERC721Received} function occur before state updates, exposing the contract to reentrancy attacks. 
    
    Regarding Weak Auth Validation and Loose Perm Management, the detector identified 15 and 12 verified instances, respectively. While maintaining high precision of 93.75\% and 100\%, the recall rates 65.22\% and 54.55\% reflect the complexity of covering all logic variations. The detected issues typically involve incorrect variable operations or omitted validation checks. Although some cases, such as redundant self-authorization for token owners in \texttt{approve} functions, may not be immediately exploitable, they introduce unnecessary complexity and latent security risks.

    The analyzer's reliance on rigid pattern matching can lead to false negatives, as it struggles to cover the diverse implementations of permission control. Additionally, the tool faces challenges in inter-procedural analysis, often overlooking logic delegated to sub-functions with diverse deployments. Furthermore, runtime-dependent parameters, such as dynamically generated address lists, cannot be precisely resolved by static analysis. These uncertainties hinder the detector from capturing all execution contexts, resulting in missed detections and occasional false alarms.

    \subsection*{RQ2: Comparison}
    To answer RQ2, we collected a range of smart contract detection tools from well-known journals and conferences in the fields of software engineering and blockchain security, such as International Symposium on Software Testing and Analysis (ISSTA), International Conference on Software Engineering (ICSE), and IEEE Transactions on Software Engineering (TSE), as well as Mythril \cite{mythril2024}, which is recommended by the official Ethereum community. These tools cover a variety of strategies for detecting vulnerabilities, including static analysis, symbolic execution, fuzz testing, and similarity detection.
    
    After the screening, we conducted experiments on the verification dataset using five detection tools: Solhint \cite{solhint2024}, NFTGuard \cite{yang2023definition}, Achecker \cite{ghaleb2023achecker}, Mythril \cite{mythril2024}, and SmartEmbed \cite{gao2020checking}. To evaluate whether these tools could accurately detect relevant defects, we randomly selected 15 contracts from the validation dataset that were confirmed to contain a specific single type of permission control vulnerability in RQ1 as the test cases for comparison—five samples per vulnerability type. And performed small-scale experiments with the five aforementioned tools. 

    We categorize the detection results into five types (i.e., \textbf{Y}, \textbf{N}, \textbf{X}, \textbf{E}, and \textbf{T}). Both \textbf{Y} and \textbf{N} indicate that the detection tool identified a vulnerability in the contract. Specifically, \textbf{Y} signifies that the vulnerability relates to permission control, while \textbf{N} indicates irrelevance. The types \textbf{X}, \textbf{E}, and \textbf{T} indicate no vulnerabilities found: \textbf{X} means successful detection, \textbf{E} signifies a detection failure, and \textbf{T} represents a timeout. 
    Additionally, the terms \textbf{ME}, \textbf{VE}, \textbf{CE}, and \textbf{LE} refer to specific errors: \textbf{ME} stands for Memory Error, \textbf{CE} denotes Compilation Error, \textbf{VE} means Value Error, and \textbf{LE} indicates Logic Error. During the experiment, we set the maximum memory limit to 24GB and the timeout threshold to 7,200 seconds.

    \begin{table}[ht]
    \small
    \centering
    \caption{Comparative Detection Results on Fifteen Smart Contracts across Analysis Tools.}
    \label{tab:comparison_results}
    \begin{tabular}{>{\centering\arraybackslash}m{3cm} >{\centering\arraybackslash}m{1.2cm} >{\centering\arraybackslash}m{1.2cm} >{\centering\arraybackslash}m{1.2cm} >{\centering\arraybackslash}m{1.2cm} >{\centering\arraybackslash}m{1.2cm} } % 使用 p{width} 设定列宽
        \toprule
        \textbf{Defect Category} & \textbf{Solhint} & \textbf{AChecker} & \textbf{NFTG} & \textbf{Mythril} & \textbf{SmartEmb}\\
        \midrule
        \multirow{5}{1.2cm}{\centering Weak Auth Validation} & N & ME & CE & X & X\\
        & N & ME & CE & X & X\\
        & N & ME & CE & X & X\\
        & N & ME & CE & X & X\\
        & N & X & CE & X & X\\
        \midrule
        \multirow{5}{1.2cm}{\centering Loose Perm Manage} & N & ME & LE & E & X\\
        & N & VE & LE & T & X\\
        & N & ME & LE & T & X\\
        & N & ME & CE & X & X\\
        & N & CE & LE & X & X\\
        \midrule
        \multirow{5}{1.2cm}{\centering Bypass Auth Reentrancy} & N & ME & X & X & X\\
        & N & CE & X & X & X\\
        & N & CE & X & E & X\\
        & N & CE & X & X & X\\
        & N & CE & X & X & X\\
        \bottomrule
    \end{tabular}
 
    \end{table}

    The experimental results on 15 vulnerable contracts are shown in Table \ref{tab:comparison_results}. Among the five contract analysis frameworks, most failed to identify and report vulnerabilities related to permission control, and some encountered significant errors during analysis. 
    Based on our analysis, the possible reasons for this are as follows:

    \noindent\textbf{Solhint:} Solhint analyzes the AST and applies a rule-based engine primarily designed for rapid code style and formatting checks. However, it lacks in-depth contract security analysis, particularly in detecting permission control vulnerabilities, as it does not perform deeper semantic or logical reasoning about contract behavior.
    
    \noindent\textbf{AChecker:} AChecker combines static analysis and symbolic execution to analyze access control in contracts, but it consumes a large amount of memory to store branching paths. Additionally, AChecker appears to have insufficient error handling, and its analysis occasionally terminates unexpectedly, potentially due to memory constraints.
    
    \noindent\textbf{NFTGuard:} NFTGuard focuses on discovering security vulnerabilities in NFT contracts through symbolic execution. However, compatibility issues may arise in practice, especially when compiler versions are inconsistent, leading to compilation failures or inconsistent analysis results. Furthermore, the tool's exclusive reliance on bytecode-based function analysis imposes substantial constraints on its applicability in large-scale automated vulnerability detection.

    \noindent\textbf{Mythril:} Mythril integrates multiple analysis modules and dynamically switches heuristic rules, theoretically allowing for in-depth detection. However, it appears to focus on analyzing control flow and variable mutations, while paying less attention to application-level logic (e.g., improperly implemented functions), which limits its effectiveness in identifying permission-related issues.
    
    \noindent\textbf{SmartEmbed:} As a similarity comparison tool, SmartEmbed performs well in certain areas. However, it may be limited by a lack of representative data samples for specific vulnerabilities, making it challenging to customize detection strategies accordingly.

\begin{figure}[htbp]
        \centering
        \includegraphics[width=0.5\textwidth]{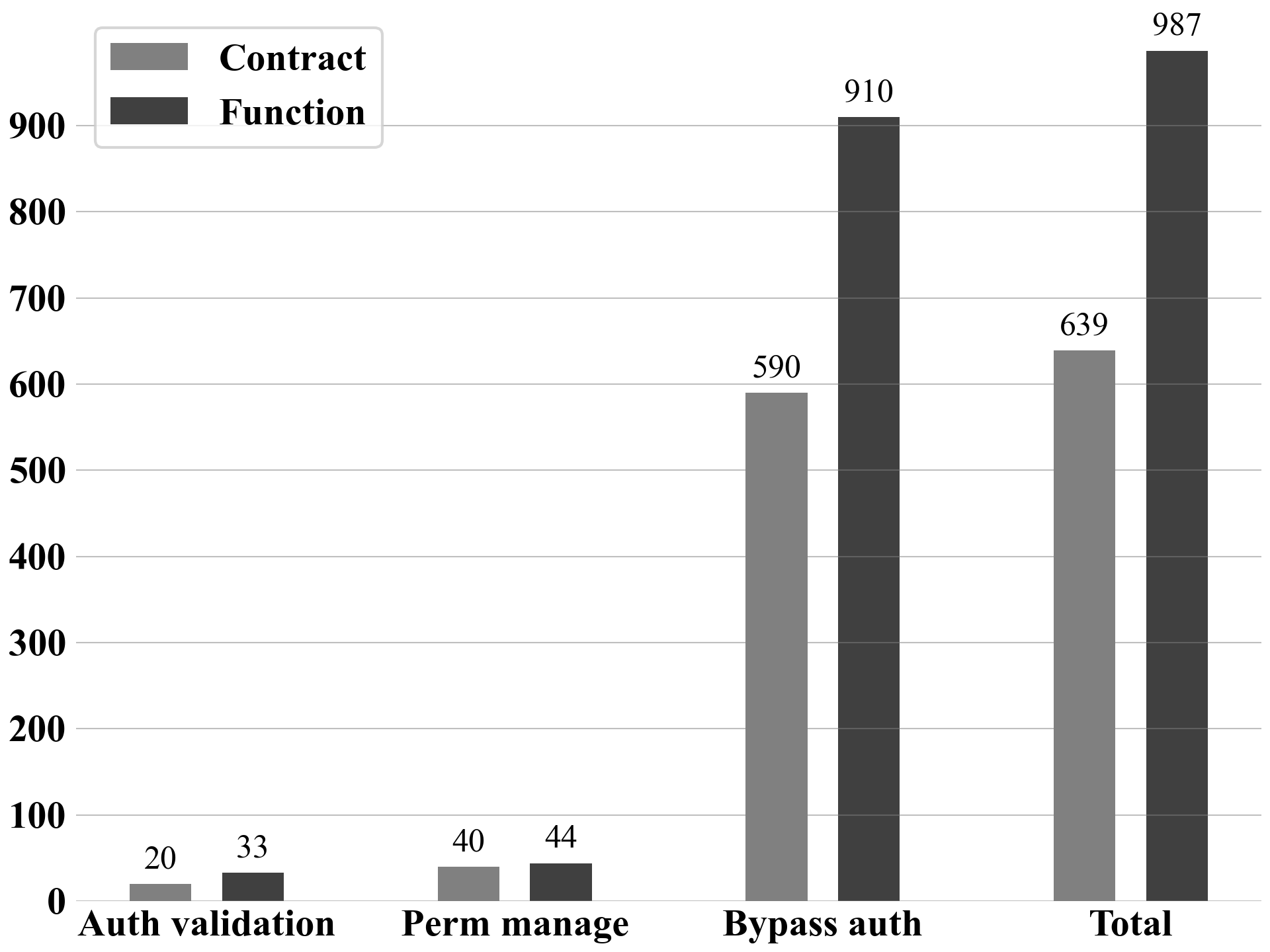}
        \caption{Distribution of Defects in the Embedding Dataset Divided into Contract and Function-Level Defect Counts (Total Represents the Union of all Three Defect Categories).}
        \label{fig:defects_in_initial_dataset}
    \end{figure}

    NFTDELTA can accelerate the analysis efficiency of contract detection tools above by leveraging preprocessing and feature analysis, eliminating the need for computationally intensive techniques.
    During the embedding phase, among the 3,146 smart contracts analysed, 639 vulnerable contracts were successfully processed and stored in the database, as illustrated in Figure \ref{fig:defects_in_initial_dataset} (For clarity, we excluded the two contracts with multiple vulnerability types). The total time spent compiling, detecting, embedding, fusing, and storing these contracts was 2,856 seconds (as recorded by our in-program timer), averaging 4.47 seconds per contract. 
    
    In the detection phase, the average time to analyze a single contract, including compilation, embedding, fusion, and storage, was 3.36 seconds, totaling 2,671 seconds across 795 contracts. NFTDELTA's average memory usage in the two phases is about 3-4 GB, highlighting its suitability for deployment in resource-constrained environments.
    It is worth noting that the reported detection times were measured using the framework's built-in timers. Due to the efficient static analysis capabilities, the time difference between the embedding and detection phases is negligible. However, differences in detection strategies for new vulnerabilities—such as pattern matching or variable tracking—and the underlying analysis techniques (e.g., static analysis vs. fuzz testing) may lead to variations in efficiency. These variations can affect the embedding process in terms of both speed and accuracy. In contrast, similarity-based analysis is independent of the embedding phase and remains unaffected by these factors.

    \subsection*{RQ3: Ablation}
    \label{subsec::RQ3_ablation}

    To answer RQ3, we use the defective contract data from the embedding dataset in \cref{sec::data_collection} and perform contract similarity detection on contracts in the validation dataset. We will evaluate NFTDELTA's similarity detection accuracy through ablation experiments. The experiment is divided into two parts: the first part verifies the impact of different similarity thresholds on the experimental results by adjusting parameters; the second part disables different modules and evaluates the contribution of each module to the detection accuracy.

    \textbf{Threshold Ablation:}
    Table \ref{tab:threshold_ablation} presents the average accuracy of the NFTDELTA framework for detecting three types of permission control vulnerabilities using the similarity analysis method across different thresholds. These results indicate that setting a lower threshold improves the model's accuracy in identifying vulnerabilities. However, even with strict evaluation criteria and analysis conducted at lower similarity thresholds, our method still exhibits a higher false-positive rate than static-analysis-based detection. 

    \begin{table}[ht]
    \small
    \centering
    \caption{Precision of Similarity Detection at Different Thresholds.}
    \label{tab:threshold_ablation}
    \begin{tabular}{>{\centering\arraybackslash}m{1.2cm} >{\centering\arraybackslash}m{0.8cm} >{\centering\arraybackslash}m{0.8cm} >{\centering\arraybackslash}m{1cm}}
        \toprule
        \textbf{Threshold} & \textbf{\#TP} & \textbf{\#FP} & \textbf{Prec(\%)}\\
        \midrule
        0.01 & 28 & 41 & 40.58\\
        \midrule
        0.1 & 29 & 46 & 38.67\\
        \midrule
        1 & 29 & 48 & 37.66\\
        \midrule
        2 & 33 & 55 & 37.50\\
        \bottomrule
    \end{tabular}
    \end{table}

    The result is likely due to the strict criteria employed in the similarity analysis method. Under the current configuration, two functions are deemed similar only if all their basic block vectors are highly analogous and the function names match the database labels exactly. While this approach minimizes false positives, it introduces new challenges. 
    With a low threshold, the model requires that all pairs of basic blocks from two functions be highly similar to identify vulnerabilities. However, inherent differences in real-world functions make such a strict criterion prone to false negatives. Conversely, using a higher threshold reduces the model's ability to detect subtle differences between similar functions, significantly increasing false positives. Therefore, based on the detection results, we consider a similarity threshold of 0.1 to be the most reliable.

    \textbf{Model Ablation:}
    We designed ablation experiments to thoroughly analyze the contribution of each module to the framework's overall performance. In these experiments, we individually disabled the sequence and the graph embedding module (Performer and FastGAT) and disabled both modules simultaneously to construct variant frameworks. Subsequently, we compared the vulnerability detection accuracy of these three variant frameworks with the complete framework (with no modules disabled). Based on the Threshold Ablation experiment results, we set the Model Ablation threshold to 0.1.

    \begin{table}[ht]
    \small
    \centering
    \vspace{-1ex}
    \caption{Ablation Experiments by Removing Different Modules and Evaluating Precision (SE Denotes Sequence Embedding and GE Denotes Graph Embedding).}
    \vspace{-1ex}
    \label{tab:model_ablation}
    \begin{tabular}{>{\centering\arraybackslash}m{1.6cm} >{\centering\arraybackslash}m{1.2cm} >{\centering\arraybackslash}m{1.2cm} >{\centering\arraybackslash}m{1.2cm} >{\centering\arraybackslash}m{1.2cm}} % 使用 p{width} 设定列宽
        \toprule
        \textbf{Variant} & \textbf{\#TP} & \textbf{\#FP} & \textbf{Prec(\%)} & \textbf{Loss(\%)}\\
        \midrule
        - SE & 146 & 258 & 36.14 & 10.94\\
        \midrule
        - GE & 36 & 77 & 31.86 & 21.49\\
        \midrule
        - (SE \& GE) & 0 & 0 & 0 & 100\\
        \bottomrule
    \end{tabular}
    \vspace{-2ex}
    \end{table}

    Table \ref{tab:model_ablation} presents the ablation results for disabling different modules, demonstrating that both the sequence and graph embedding modules contribute positively to the model's overall performance.
    Disabling the sequence embedding module will significantly lose semantic information, causing the model to become overly permissive in its vulnerability assessments. This results in increased false positives due to overly optimistic predictions.
    Conversely, disabling the graph embedding module results in a substantial decline in accuracy, underscoring its essential role in capturing global structural patterns.
    When both modules are disabled, the model is forced to rely solely on low-level instruction semantics, which are insufficient for modeling the full contextual relationships within a contract, rendering the model incapable of providing helpful information under strict similarity evaluation criteria.

\section{RELATED WORK}
\label{sec::related}

\subsection{Permission Control Vulnerability}
    Numerous studies have focused on detecting permission control vulnerabilities in smart contracts \cite{yang2024uncover} \cite{liu2020smacs} \cite{liu2022finding}  \cite{luu2016making} \cite{yang2023definition} \cite{zhong2024prettysmart} \cite{liang2024identity}.
    Achecker, proposed by Ghaleb et al. \cite{ghaleb2023achecker}, employs a bytecode-based static data-flow analysis to systematically identify access control checks and their associated state variables. It employs taint analysis to prune infeasible paths and examine whether attackers can manipulate critical access-control instructions and state variables. To further reduce false positives, Achecker applies symbolic execution to filter out potentially intended program behaviors.
    Xiao et al. \cite{xiao2025wakemint} introduced WakeMint, a symbolic execution-based framework for detecting sleep-minting vulnerabilities. WakeMint implements specialized detectors for four categories of sleep-minting attacks by extracting key information about token transfers and leveraging constraint-solving to determine whether functions exhibit sleep-minting characteristics. In an evaluation conducted on 11,161 real-world smart contracts, WakeMint achieved a high detection accuracy of 87.8\%.

\subsection{Similarity Detection}
    Similarity detection, also known as metric-based techniques, initially gained prominence in code clone detection (CCD) research \cite{he2020characterizing} \cite{ain2019systematic}. 
    In recent years, this approach has been increasingly utilized in code similarity analysis and vulnerability detection for smart contracts \cite{zhang2024combining} \cite{zhu2022bytecode} \cite{tian2022ethereum}. 
    Gao et al. \cite{gao2020checking} introduced an automated method that employs word embedding and vector space comparison to learn the characteristics of smart contracts. This method parses contracts into word streams that capture code structure information. It converts code statements and functions into vectors that encode both syntax and semantics, allowing the identification of potential problems by comparing vector similarities with known vulnerable code.
    Huang et al. \cite{huang2021hunting} proposed a similarity comparison detection method for contract bytecode. They standardized the bytecodes generated by different compilers at both the data and instruction levels, filtering out noise code irrelevant to the contract by simulating bytecode execution. Additionally, they developed an unsupervised graph embedding algorithm that converts the code graph into a quantifiable vector, enabling the identification of potentially vulnerable contracts by comparing their similarities with known vulnerable contracts.

\section{CONCLUSION}
\label{sec::conclusion}

This paper defines three common permission-control vulnerabilities in NFT smart contracts and develops static analysis detectors to identify them. We also introduce an efficient vulnerability detection tool, NFTDELTA, which is based on a multi-view feature fusion method that integrates sequence and graph features of contract functions. We applied NFTDELTA to several popular NFT contracts. As a result, NFTDELTA achieved an average detection accuracy of 97.92\% and an F1 score of 81.09\% across three vulnerability types. Its similarity comparison module demonstrated high efficiency and strong detection capabilities, providing robust security assurance for NFT contracts and offering valuable support to users and developers. 

However, we acknowledge that NFTDELTA's similarity detection still has room for improvement. In future work, we plan to introduce training-based methods to improve the model's semantic understanding, refine the embedding and similarity computation processes, and expand the dataset to improve the generalizability of detection results. These improvements are expected to increase detection accuracy and further reinforce NFTDELTA's reliability in real-world vulnerability analysis.

\bibliographystyle{ACM-Reference-Format}
\bibliography{base}

\end{document}